\documentclass[doublecol]{epl2}
\usepackage{graphicx}
\usepackage{comment}
\usepackage{bm}
\usepackage{ifpdf}
\usepackage{floatrow}
\usepackage{amssymb}
\usepackage{amsmath}
\usepackage{aas_macros}
\usepackage{floatrow}
\usepackage{makecell}
\usepackage{caption}
\usepackage{stackrel}
\usepackage{subcaption}

\title{Tunneling time and Faraday/Kerr effects in 
$\mathcal{PT}$-symmetric systems}

\author{Vladimir~Gasparian\inst{1} \and Peng~Guo\inst{2,1,3}, Antonio  P\'erez-Garrido\inst{4} \and Esther~J\'odar\inst{4}}
\shortauthor{V.~Gasparian \etal}

\institute{\inst{1}Department of Physics and Engineering,  California State University, Bakersfield, CA 93311, USA\\
\inst{2}College of Arts and Sciences,  Dakota State University, Madison, SD 57042, USA\\
\inst{3}Kavli Institute for Theoretical Physics, University of California, Santa Barbara, CA 93106, USA\\
\inst{4}Departamento de F\'\i sica Aplicada. Hospital de Marina, Universidad Politécnica de Cartagena (UPCT), 30202 Cartagena, Murcia (Spain)}

\abstract{
We review the generalization of tunneling time and anomalous behaviour of Faraday and Kerr rotation angles  in parity and time ($\mathcal{P}\mathcal{T}$)-symmetric systems. Similarities of two phenomena are  discussed, both exhibit a phase transition-like anomalous behaviour in certain range of model parameters. Anomalous behaviour of  tunneling time and   Faraday/Kerr   angles  in  $\mathcal{P}\mathcal{T}$-symmetric systems  is caused by the motion of poles of scattering amplitudes in energy/frequency complex plane.
}

\begin{document}

\maketitle

\section{Introduction}

 During the past two decades, parity-time ($\mathcal{PT}$)-symmetric Hamiltonians are studied in various areas of physics such as optics \cite{Ruschhaupt_2005,2005-1,2011-1,2012-1,Longhi_2017},  quantum mechanics  \cite{muga1,mos1,tb1}, and classical wave systems  \cite{son1,sound1}. For example, in optics, $\mathcal{PT}$-symmetric arrangements of gain and loss media have been studied, where the gain balances the loss, leading to interesting phenomena such as unidirectional invisibility and anomalous point oscillations  \cite{unid1,unid2}.  Note that although $\mathcal{PT}$-symmetric systems have been extensively studied and have theoretical and experimental support, their implementation in certain physical systems can still be difficult. Nevertheless, they remain an active research field with great potential for new discoveries and technical applications in optical isolators, sensors, magnetic storage,
 magneto-optical modulators, magneto-optical switches, and magneto-optical circulators.

In this  letter, we present a brief review on  (i) the generalization of the concept of tunneling time in $\mathcal{PT}$ symmetric systems and (ii) some anomalous behaviours of magneto-optic effects in $\mathcal{PT}$-symmetric systems. Both  quantum tunneling time and magneto-optic effects are the consequence of  propagation of either quantum wave or optical wave through barriers, hence both phenomena are   closely related to the transmission and reflection amplitudes of scattering of a quantum particle or light off barriers, and can be described in a similar framework.

 In standard (Hermitian) quantum mechanics,    both tunneling time of a quantum particle and Faraday and Kerr rotations of a electromagnetic wave through real potential barriers   are not new subjects, and  a wide variety of theoretical and experimental work on both topics have been carried out extensively in the past:

(i) Tunneling time: Substantial research has been conducted on the tunneling time problem, see e.g. Refs.~\cite{landauer1994,fayer2001,gasparian2000} and references therein). This area of exploration is particularly focused on nanostructures and mesoscopic systems with sizes smaller than 10nm. In such systems, the tunneling time assumes significance as it becomes a key factor in determining various transport properties. Notably, it plays a vital role in phenomena such as the frequency-dependent conductivity response of mesoscopic conductors~\cite{buttiker1994} and the occurrence of adiabatic charge transport~\cite{brouwer1998,zhou1999}.
  More recently, 
another kind of problems have arisen in ultrafast science or in attosecond physics (e.g., the investigation of electron correlation effects, photoemission delay, ionization tunneling, etc), where tunneling time experiments play an important role
as unique and
powerful tool that allows in electronic monitoring with subatomic resolution both in space and time. 
The measurement of tunneling time in attosecond experiments (attosecond=$10^{-18}s$)
offers a fruitful opportunity to understand the role of time in quantum mechanics, which has been controversial since the appearance of quantum mechanics, see, e.g., Refs. \cite{gasparian2000,MUGA2004357,landauer1994}.

(ii) Magneto-optic effects: 
The development of electromagnetic theory and atomic physics has been largely influenced by the study of the magneto-optic effects as Faraday rotation (FR)  and Kerr rotation (KR). In these magneto-optical phenomena, an electromagnetic  wave propagates through a medium altered by the presence of an external magnetic field. In such magneto-optical materials (also referred as gyrotropic or gyromagnetic), left- and right-rotating elliptical polarizations propagating at different speeds result in  a rotation of the planes of the transmitted (FR) and reflected (KR) light.  FR and KR effects are essential for optical communication technologies \cite{1982-1}, optical amplifiers \cite{2012-1,2000-1}, and photonic crystals \cite{prl-1,2009-N}. In addition, the KR is also an extremely accurate and versatile research tool and can be used to determine quantities,  such as anisotropy constants, exchange-coupling strengths and Curie temperatures (see, e.g. \cite{kr}).

\section{General theory of tunneling time and magneto-optic effects in  $\mathcal{PT}$-symmetric systems}
A brief summary of general theory of tunneling time and magneto-optic effects in  $\mathcal{PT}$-symmetric systems is given in this section, more details can be found in \cite{PhysRevResearch.4.023083,Gasparian:2022acf,Guo:2022jyk,Perez-Garrido:2023pfg}.

\subsection{Tunneling time}

 The concept of tunneling or delay time for a quantum particle tunneling through real potential barriers is conventionally defined through the integrated density of states, which is proportional to the imaginary part of full Green's function of systems and positive definite in a real potential scattering theory.   Following the definition in Refs.~\cite{PhysRevB.47.2038,PhysRevB.51.6743,PhysRevA.54.4022},  two components of the traversal time  $\tau_E$ can be introduced by  ($\hbar=1$)
 \begin{equation}
  \tau_E= \tau_2  + i \tau_1  = - \int_{ -  \frac{\Lambda}{2}}^{ \frac{\Lambda}{2} } d x \langle x | \hat{G} (E) | x \rangle , \label{taudef}
 \end{equation}
where $\tau_1$ and $\tau_2$ represent  B\"uttiker-Landauer  tunneling time   and the Landauer resistance  respectively.   The $\Lambda$ stands for the    length of potential barrier, and  the $\hat{G} (E)  =  \frac{1}{ E- \hat{H} } $ is the Green's function operator of system. 
 As shown in Refs.~\cite{PhysRevB.51.6743,PhysRevA.54.4022},  
 two components of the traversal time $\tau_E$ are linked to the scattering and transport amplitudes explicitly by   
  \begin{equation}
  \tau_2  + i \tau_1   = \frac{d  \ln \left [t (k)  \right ]}{d E}+ \frac{ r^{(l)} (k) + r^{(r)} (k) }{4 E}  , \label{integGtr}
 \end{equation}
  where  $t(k)$ and $r^{(l/r)}(k)$ are the transmission and left/right reflection amplitudes respectively, and $k = \sqrt{2 m E}$ is the momentum of particle.   The transmission and reflection amplitudes can be obtained by finding scattering solutions of Schr\"odinger equation,  
  \begin{equation}
\hat{H} |  \Psi_{E} \rangle = E |  \Psi_{E} \rangle, \ \ \ \  \hat{H} =  - \frac{1}{2m } \frac{d^2}{ d x^2} +V(x),
 \end{equation}
where $m$ denotes the mass of particle, and $V(x)$ is the interaction potential.

  In the conventional real potential scattering theory,    the development of concept of tunneling or delay time     is  fundamentally  based  on   counting the probability that a particle  spends  inside of a barrier, see  e.g. Refs.~\cite{PhysRev.98.145,PhysRev.118.349,goldberger2004collision}. However, in complex potential scattering theory, the norm of states is no longer conserved,    the probability interpretation of tunneling time becomes problematic. This can be understood by examining     the spectral representation of Green's function in a complex potential scattering theory,  which now depends on   the eigenstates of both $\hat{H} $ and its adjoint $\hat{H}^\dag$,
   \begin{equation}
  \hat{G} (E)   = \sum_{i} \frac{ |  \Psi_{E_i} \rangle \langle \widetilde{\Psi}_{E_i } |}{E- E_i},
 \end{equation}  
  where
   \begin{equation}
\hat{H}^\dag |  \widetilde{\Psi}_{E} \rangle = E |  \widetilde{\Psi}_{E} \rangle.
 \end{equation}
   In general the discontinuity of Green’s function crossing the branch cut in complex E-plane is a complex function. However, see Ref.~\cite{PhysRevResearch.4.023083}, it is real under PT symmetry and equal to the imaginary part of the Green’s function, though it may not always be positive as a consequence of norm violation due to the complex potential.   Hence, the conventional definition of density of  states gets lost, and one should generalize and redefine correctly the tunneling time. In this review letter the tunneling time through $\mathcal{P}\mathcal{T}$-symmetric  barriers is  defined by Eq.~(\ref{taudef}). The generalized density of  states in such a  system is taken as the imaginary part of Green's function, and  $\tau_1$  now may be interpreted as a  generalized   B\"uttiker-Landauer  tunneling time. The sign of generalized tunneling time $\tau_1$ is directly related to the potential barriers, that either tend to keep a particle in or force it out. When $\tau_1$ is negative it behaves similar to a forbidden gap in a periodic system, being unreachable.

\begin{figure}
\begin{center}
\includegraphics[width=0.5\textwidth]{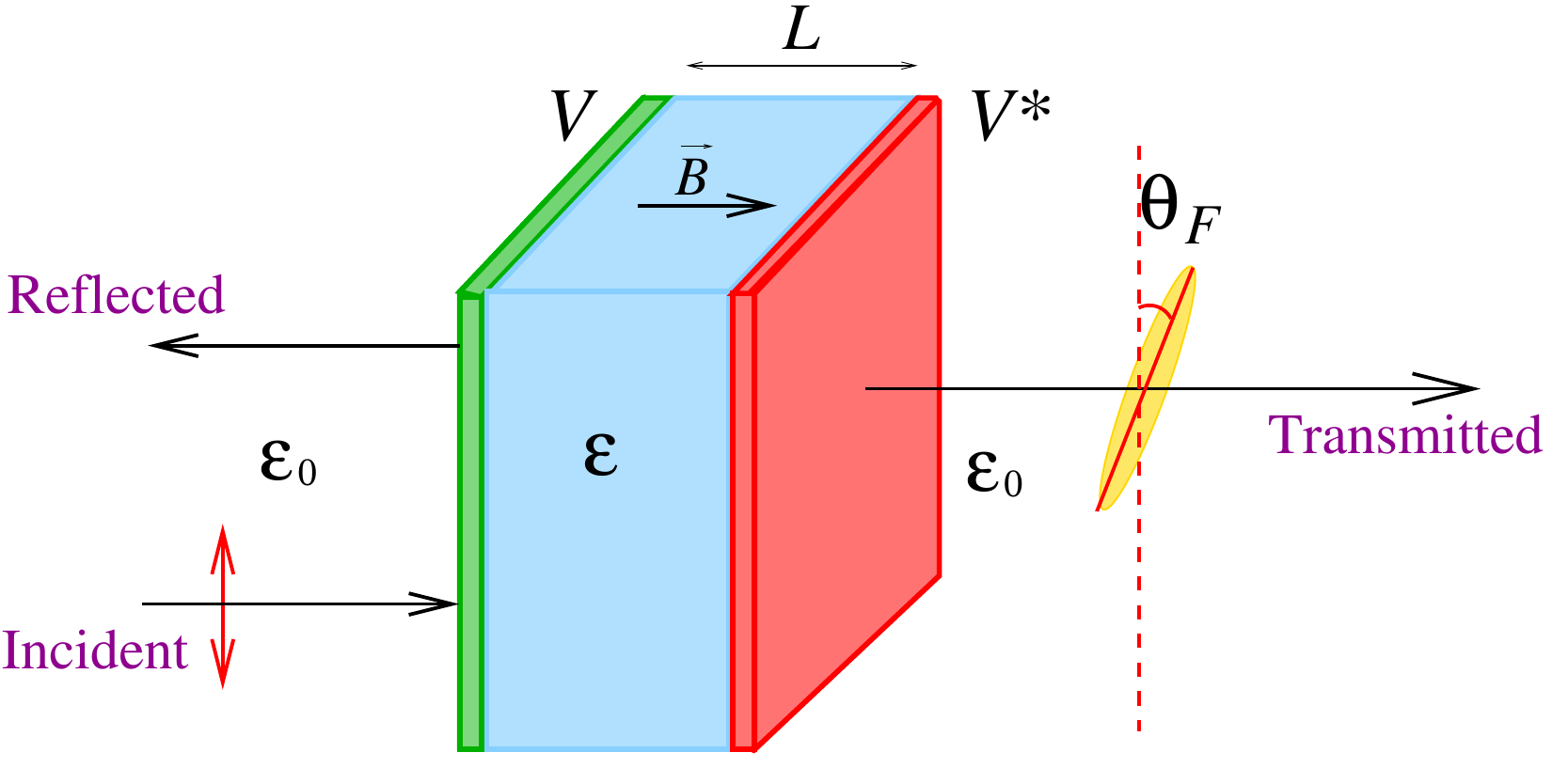}
\caption{ Demo plot of  a $\mathcal{P}\mathcal{T}$-symmetric dielectric slab model with two balanced complex narrow slabs   placed at  both ends of a real dielectric  slab.  }\label{slabplot}
\end{center}
\end{figure}

\subsection{Magneto-optic effects}

We consider an incident linearly polarized electromagnetic plane wave with an angular frequency $\omega$ entering the system from the left, propagating along the $x$ direction. The electric field's polarization direction of the incident wave is aligned with the z-axis: $\boldsymbol{E}_0(x) = e^{i \frac{\omega}{c}\sqrt{\epsilon_0}  x} \hat{z}$, where 
$\epsilon_0$ denotes the dielectric constant of vacuum.  A magnetic field  $\boldsymbol{B}$   is applied in the $x$-direction, as depicted in Fig.~\ref{slabplot}. The scattering of EM wave is described by, see e.g. Refs.~\cite{Josh,PhysRevLett.75.2312},
\begin{equation}
\left[\frac{d^2}{dx^2} + \frac{\omega^2 \epsilon_{\pm}(x)}{c^2}\right]E_{\pm}(x) = 0, \label{maxwelleq}
\end{equation}
where $E_{\pm} = E_y \pm iE_z$ represents the circularly polarized electric fields. The variable $\epsilon_{\pm}(x)$ is defined as follows:
\begin{equation}
\epsilon_{\pm}(x) =
\begin{cases}
\epsilon +V (x) \pm g, & \text{if } x \in [- \frac{L}{2}, \frac{L}{2}], \\
\epsilon_0, & \text{otherwise},
\end{cases}
\end{equation}
where  $L$   represents the   length of   dielectric slab and $\epsilon$ is positive and real permittivity of the slab, see  Fig.~\ref{slabplot}. The $V(x)$ denotes the additional potential barriers that are placed inside of the slab, which can be easily  manipulated and adjusted to implement $\mathcal{P}\mathcal{T}$ symmetry requirement.  The $g$ is the gyrotropic vector along the magnetic-field direction. The external magnetic
field $\boldsymbol{B}$ is included into the gyrotropic vector $g$ to make the calculations valid for the cases of both external magnetic
fields and magneto-optic materials.
The magnetic field causes the direction of linear polarization of  both transmitted and reflected wave to rotate. As a consequence, both  the outgoing transmitted and reflected waves exhibits elliptical polarization. The major axis of the ellipse is rotated relative to the original polarization direction. The real part of the rotation angle describes the change in polarization for linearly polarized light, while the imaginary part indicates the ellipticity of the transmitted or reflected light.

The complex rotational parameters characterizing the transmitted light 
can be expressed  in terms of transmission amplitudes by
\begin{equation}
 \theta^{T}_2 + i \theta^{T}_1 =   \frac{1}{2}\ln\frac{t_+ (\omega)}{t_- (\omega)}, \label{FR1}
\end{equation}
where $t_\pm$  represent the transmission  amplitudes  of transmitted electric fields. In the case of a weak magnetic field ($g\ll 1$), a perturbation expansion can be applied. The leading-order contribution can be obtained by expanding   $t_\pm$ around the refractive index of the slab in the absence of the magnetic field $\boldsymbol{B}$:
\begin{equation}
\theta^{T}_{2}  + i \theta^{T}_1 =   \frac{g}{2n}\frac{\partial\ln  \left [ t (\omega) \right ] }{\partial n}, \label{1mb}
\end{equation}
where, $n = \sqrt{\epsilon}$ represents the refractive index of the slab. Similarly, the leading-order expressions  of complex angles of Kerr rotation,  in the case of a weak magnetic field, are given by:
\begin{equation}
\theta^{R^{(l/r)}}_{2} + i \theta^{R^{(l/r)}}_1 =  \frac{g}{2n}\frac{\partial\ln  \left [r^{(l/r) }  (\omega) \right  ] }{\partial n}, \label{ref}
\end{equation}
where $r^{(l/r)} $ is  the left/right  reflection   amplitudes in the absence  of magnetic field $\boldsymbol{B}$.

\subsection{$\mathcal{PT}$ symmetry constraints on scattering amplitudes}

The $\mathcal{PT}$ symmetry can be implemented in quantum tunneling time and  magneto-optic effects by imposing conditions on interaction potential in   Schr\"odinger equation   and on dielectric permittivity in Eq.(\ref{maxwelleq}): $V(x ) = V^*(-x)$.   As discussed in Ref.~\cite{PhysRevResearch.4.023083}, the parametrization of scattering matrix only requires three independent real functions in a $\mathcal{P}\mathcal{T}$-symmetric system: one inelasticity, $\eta \in [1, \infty]$, and two phaseshifts, $\delta_{1,2}$.  In terms of $\eta $ and  $\delta_{1,2}$,  the transmission  and reflection amplitudes are given by
 \begin{align}
 t   &=\eta  \frac{ e^{2 i \delta_1 }+ e^{2 i \delta_2 }}{2},  \nonumber \\
    r^{(r/l)} &= \eta \frac{ e^{2 i \delta_1 }-   e^{2 i \delta_2 }}{2} \pm i \sqrt{\eta^2-1} e^{i (\delta_1+ \delta_2)}.
 \end{align}

As the consequence of $\mathcal{PT}$ symmetry constraints,  two components of the traversal time  are also given  in terms of $\eta $ and  $\delta_{1,2}$ by
   \begin{align}
  \tau_1    & =  \frac{d  ( \delta_1 + \delta_2 )  }{d E}  +  \eta  \frac{    \sin (2 \delta_1 )  - \sin (2 \delta_2 )     }{4 E}  , \nonumber \\
  \tau_2     & =  \frac{d   \ln \left [ \eta  \cos ( \delta_1 - \delta_2 ) \right ]  }{d E}  +\eta  \frac{    \cos^2  \delta_1 - \cos^2  \delta_2   }{2 E}  .
 \end{align}
Only three FR and KR angles are independent,  the  FR and KR angles are given  in terms of $\eta $ and  $\delta_{1,2}$  by
\begin{align}
\theta^{T}_1  &= \theta^{R}_1  =  \frac{g}{2n} \frac{\partial (  \delta_1 +  \delta_2 )}{\partial n}, \ \ \ \
\theta^{T}_{2}  =
 \frac{g}{2 n} \frac{\partial  \ln \left [ \eta \cos ( \delta_1 - \delta_2) \right ]}{\partial n}   ,   \nonumber \\
  & \theta^{R^{(r/l)}}_{2}  =
 \frac{g}{2 n} \frac{\partial  }{\partial n} \ln  \left | \eta   \sin ( \delta_1 - \delta_2)   \pm  \sqrt{\eta^2-1}   \right |. \label{allthetas}
\end{align}
   $\theta^{R^{(r/l)}}_{2} $ and $\theta^{T}_{2}$ are   related by 
$  \theta^{R^{(r)}}_{2} + \theta^{R^{(l)}}_{2}    =     \frac{2 T }{  T  -1}  \theta^{T}_{2}  , \label{theta2RT}  $ 
  where $T    =\eta^2 \cos^2 ( \delta_1 - \delta_2)  $ denotes the transmission coefficient.

\section{A simple exact solvable  $\mathcal{PT}$-symmetric model}

A simple  $\mathcal{PT}$-symmetric contact interaction potential model is adopted in this letter  to illustrate some usual features in both tunneling time and magneto-optic effects in $\mathcal{PT}$-symmetric systems.  The $\mathcal{PT}$-symmetric interaction potential for a single  cell of barrier is chosen as
  \begin{equation}
  V(x) =  V \delta(x+\frac{L}{2}) + V^* \delta(x -\frac{L}{2}) , \ \ \ \ V = |V| e^{i \varphi_V} = V_1+ i V_2,
  \end{equation}
which  represents two complex-conjugate impurities placed inside of a single cell.  One is absorbing with loss and another is emissive with an equal amount of gain.  The closed forms of scattering solutions can be   obtained for contact interaction potential model:

(i)   Transmission and reflection amplitudes for a quantum particle tunneling through a single cell of barrier of length $\Lambda $ $ (\Lambda = L +L_0 )$ are given by
 \begin{equation}
 t_0(k)    = \frac{ \csc (k L) e^{ i k L_0}}{   \mathcal{R}(k) - i \mathcal{I} (k) }     ,   \ \  \  \ 
   r^{(l/r)}_0(k)    =i\frac{   Q^{(l/r)} (k) e^{i  k L_0}}{\mathcal{R}(k)-i \mathcal{I}(k)} , \label{t0r0expression}
\end{equation}
where $L_0$ stands for the separation between two slabs, and
  \begin{align}
&    \mathcal{R}(k)   =  \cot (k L)    +  2    \frac{  m  | V|   \cos \varphi_V   }{k  }       , \nonumber \\
&   \mathcal{I}(k)   =   1  -   2   \left  ( \frac{  m|V| }{k} \right  )^2   - 2 \left  (  \frac{  m  | V|   \cos \varphi_V  }{k  } \right  )  \cot (k L)     , \nonumber \\
& Q^{(l/r)} (k)   =-  2 \frac{ m |V|}{k}    \left [   \frac{  \cos (    k L \mp \varphi_V)  }{ \sin (k L ) }+  \frac{ m |V|}{k}       \right ] . \label{RIQexpression}
\end{align}

(ii)  The  transmission and reflection amplitudes in 
Eq.~(\ref{t0r0expression}) also apply  in the case of    Faraday and Kerr effects   for  a single cell with slab of length $\Lambda $ by replacing $k $   by $\frac{\omega n}{c}$.  The expression of functions $ {\it \mathcal{R}}(\omega) $, $ {\it \mathcal{I}}(\omega) $ and $Q^{(r/l)}(\omega) $ are given in Eq.(15) and Eq.(18) in Ref.\cite{Perez-Garrido:2023pfg}.

\begin{figure}
\begin{center}
\includegraphics[width=1\textwidth]{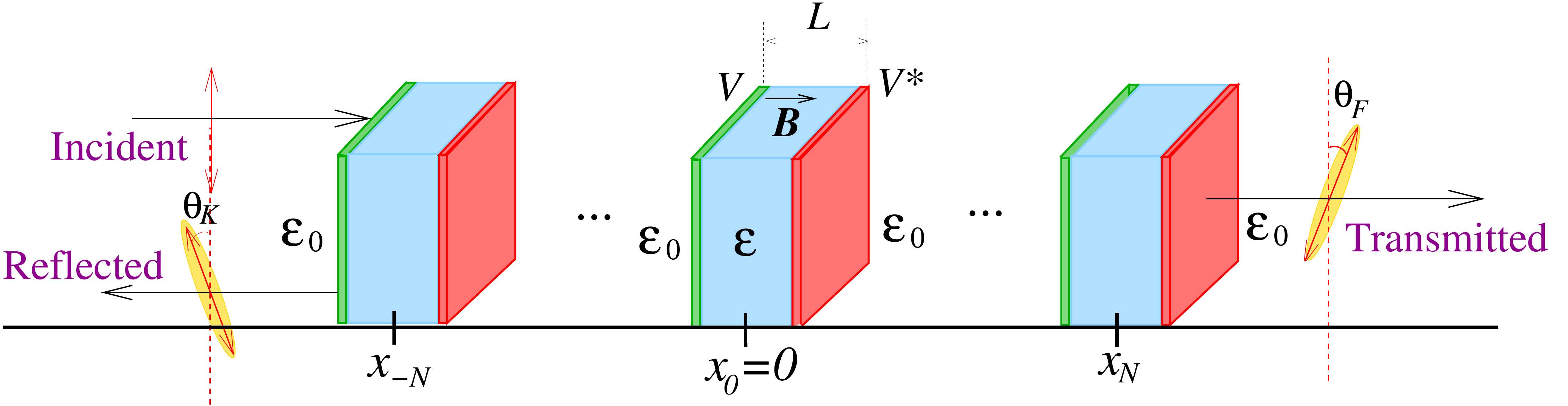}
\caption{Schematic of a one-dimension multiple cells $\mathcal{P}\mathcal{T}$-symmetric photonic heterostructure. }\label{fig.sch}
\end{center}
\end{figure}

\subsection{Periodic multiple cells $\mathcal{PT}$-symmetric systems}

It is known that when  the wave propagation through a medium is described by a differential equation of second order,
the expression
for the total transmission from the finite periodic system for any  waves (sound and electromagnetic) depends on the unit cell transmission,
the Bloch phase and the total number of cells. 
The infinite periodic $\mathcal{P}\mathcal{T}$-symmetric structures exhibit unusual properties, including 
the band structure, Bloch oscillations, unidirectional propagation and enhanced sensitivity,  see e.g., Refs.~\cite{ben,shin,Mus,mid1} and Refs.~therein.
However, the case of scattering
in a finite periodic system composed of an arbitrary number of
cells/scatters has been less investigated, despite that any open quantum system generally consists of a finite system coupled with an infinite environment. 
In Hermitian theory, the averaged  physical observables  of a finite system   approach the limit that depends on the crystal-momentum of an infinite periodic system as the size of a finite system is increased. However in  $\mathcal{P}\mathcal{T}$-symmetric systems, new challenges emerge, the large size limit of a finite size system is only well-defined conditionally.

The $\mathcal{P}\mathcal{T}$-periodic symmetric structure that consists of $2N+1$ cells, see Fig.\ref{fig.sch}, can be assembled on top of single cell.   Following Refs.~\cite{Esther-1997,Guo:2022jyk,Perez-Garrido:2023pfg},   a generic expressions for the transmission and  reflection amplitudes for  $2N+1$ cells of $\mathcal{P}\mathcal{T}$-periodic symmetric structure  in  both   tunneling time and magneto-optic effects cases can be presented as: 
 \begin{align}
& t   = \frac{ 1 }{   \cos (\beta (2N+1)  \Lambda )   + i  Im \left [ \frac{1}{t_0  }    \right ] \frac{\sin  (\beta (2N+1) \Lambda)}{\sin (\beta \Lambda)}    },  \nonumber \\
&  \frac{ r^{(l/r)}  }{t } =    \frac{ r^{(l/r)}_0    }{t_0  }  \frac{\sin (\beta (2N+1) \Lambda)}{\sin (\beta\Lambda) }     . \label{tandrexpress}
\end{align}
The $\beta$ plays the role of crystal-momentum for a periodic lattice and is related to $k$ or $\omega$ by
\begin{equation}
 \cos (\beta \Lambda ) = Re \left [ \frac{1 }{t_0 }   \right ]   . 
\end{equation}

The factor $\frac{\sin (\beta (2N+1) \Lambda)}{\sin (\beta \Lambda)}$ in both transmission and reflection amplitudes reflects the combined   interference and diffraction effects in finite periodic systems, which occur naturally in Hermitian  one-dimensional finite size periodic systems. It is interesting to see that Eq.(\ref{tandrexpress}) hold up for both Hermitian and $\mathcal{P}\mathcal{T}$-symmetric systems, which is highly non-trivial since the usual probability conservation property 
for Hermitian systems must be generalized in  $\mathcal{P}\mathcal{T}$-symmetric systems.   As pointed out in Ref.~\cite{Guo:2022jyk}, the   scattering amplitudes for a periodic multiple cells system can be   related to single cell   amplitudes in a compact fashion. 
This is ultimately due to the   factorization of dynamics living in two distinct physical scales: short-range dynamics in a single cell and long-range collective effects of  the periodic structure of entire system.    The short-range interaction dynamics   is described by single cell scattering amplitudes and  the $\beta$  represents the long range correlation effect of entire lattice system. The occurrence of factorization  of  short-range dynamics and long-range collective mode   has been known   in both condensed matter physics and nuclear/hadron physics.  As examples,  particles interacting with short-range potential in a periodic box or trap,      quantization conditions can be given in a compact formula   that is known as Korringa–Kohn–Rostoker   method  \cite{KORRINGA1947392,PhysRev.94.1111} in condensed matter physics, L\"uscher formula  \cite{Luscher:1990ux}  in lattice quantum chromodynamics  and  Busch-Englert-Rza\.zewski-Wilkens  formula \cite{Busch98}   in nuclear physics community. Other related useful discussions can be found in  e.g. Refs.~\cite{Guo_2022_JPG,PhysRevD.103.094520,Guo_2022_JPA,PhysRevC.103.064611,Guo:2023ecc}.

\subsection{Spectral singularities and their impact on tunneling time and magneto-optic effects in  $\mathcal{PT}$-symmetric systems}

\begin{figure}
\begin{center}
\includegraphics[width=0.8\textwidth]{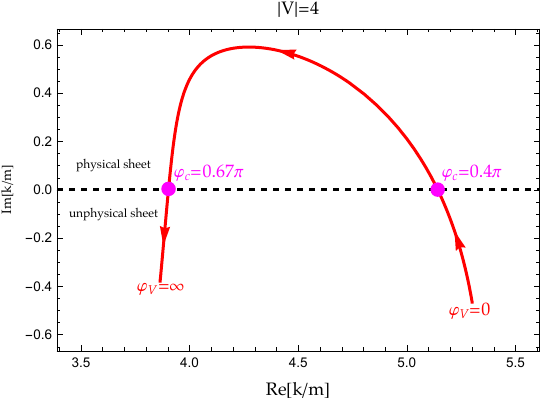}
\caption{   The motion of poles in complex $k/m$-plane as a function of increasing $\varphi_V$ for   tunneling time of a single cell defined   in Eq.(\ref{t0r0expression}) and Eq.(\ref{RIQexpression}).   The arrows  indicate increasing   $\varphi_V$ directions. The $\varphi_V$ values of  spectral singularities are indicated by $\varphi_c$'s.
The model parameters are taken as:   $ |V| =4 $ and $m L=1$.}\label{motionpolesplot}
\end{center}
\end{figure}

\begin{figure}
\begin{center}
\includegraphics[width=0.8\textwidth]{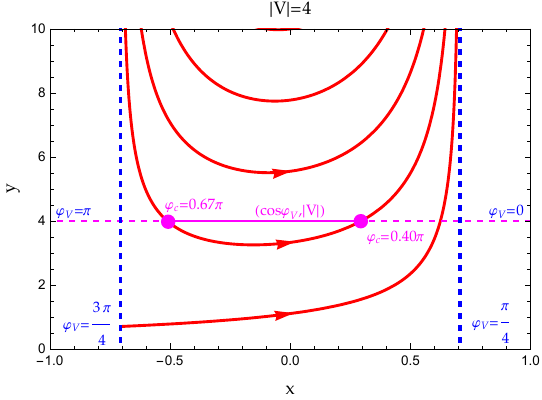}
\caption{ Spectral singularities condition plot: the parametric plot of solid red  curve is generated with $(x,y)$  coordinates    given by left-hand side of  Eq.(\ref{specsingular})     as the function of $k/m$.  The solid red  curve is bound by two blue vertical lines located at $x= \pm \frac{ 1}{ \sqrt{2 } }$.  The purple line is generated with coordinates of $(\cos \varphi_V,  |V|)$ by varying $\varphi_V$ in the range of $[0, \pi]$. The arrows  indicate increasing $k/m$   directions. The  spectral singularities for fixed $ m|V|$ are given by intersection of purple line and red curve.    The model parameters are chosen as:  $|V| =4 $, and  $mL=1$.  }\label{spectimeplot}
\end{center}
\end{figure}

 Two types of singularities are present in scattering amplitudes:  (1) a branch cut siting along the positive real axis in complex $k$ or $\omega$-plane   that separate physical sheet (the first Riemann sheet) and unphysical sheet  (the second Riemann sheet);    (2) poles of transmission and reflection amplitudes. These poles are called   spectral singularities of a non-Hermitian Hamiltonian  when they show up on real axis \cite{PhysRevLett.102.220402,Ahmed_2009,PhysRevB.80.165125},   which  yields   divergences of reflection and transmission coefficients of scattered states.

The motion of poles  in complex $k$ or $\omega$-plane, Fig.~\ref{motionpolesplot},  has some profound impact on the value of  $\tau_1$ in tunneling time  and Faraday and Kerr rotation angles $\theta_1^T$ and $\theta_1^R$ in  $\mathcal{PT}$-symmetric systems. The location of these poles are model parameters dependent and can be found by solving $1/t =0$.  The normal or anomalous behaviours of $\tau_1$ and $\theta_1^{T/R}$  are determined by the location of poles:  when the poles are all located in unphysical sheet (the second Riemann sheet), $\tau_1$ and $\theta_1^{T/R}$   remains positive. As poles moves close to and ultimately  cross real axis into physical sheet (the first Riemann sheet), the poles generate a   enhancement in $\tau_1$ and $\theta_1^{T/R}$   near the location of poles.    The spectral singularities occur  when poles are located on real axis, the transmission and reflection amplitudes diverge at location of poles. This can be easily understood   with the motion of  a single pole. Near the pole, the transmission amplitude is approximated by
\begin{equation}
t(k) \propto \frac{1}{k-k_{pole}  } = \frac{k-k_{re} - i  \gamma}{ (k-k_{re})^2+ \gamma^2}, 
\end{equation}
where  $k_{pole} = k_{re} + i \gamma$,  being $k_{re}$ and $\gamma$ the real and imaginary parts of pole position. The location of  pole   in physical sheet  or  unphysical sheet is determined by sign of $\gamma$:  unphysical sheet if $\gamma <0$ and physical sheet if $\gamma >0$. The tunneling time $\tau_1$ or Faraday/Kerr rotation angle $\theta_1^{T/R}$  near the pole is thus dominated by  
\begin{equation}
\tau_1 \sim \frac{m}{k} \frac{   \gamma}{ (k-k_{re})^2+ \gamma^2},
\end{equation}
hence as pole moves across real axis into physical sheet, $\gamma$ changes  its sign.

  \begin{figure*}
 \centering
  \begin{subfigure}[b]{0.38\textwidth}
\includegraphics[width=0.89\textwidth]{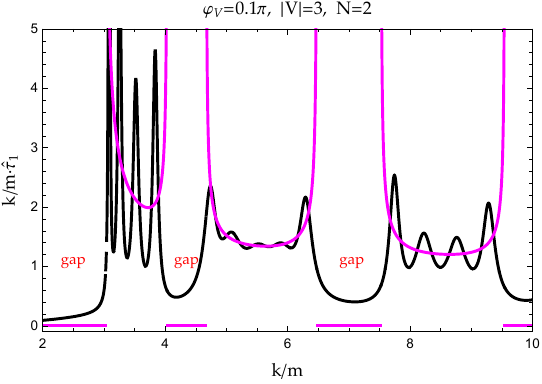}
\caption{     }\label{tau1plot1}
\end{subfigure} 
\begin{subfigure}[b]{0.38\textwidth}
\includegraphics[width=0.89\textwidth]{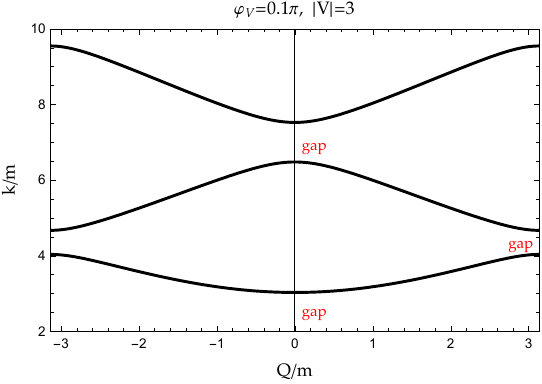}
\caption{    }\label{bandplot1}
\end{subfigure}
 \begin{subfigure}[b]{0.38\textwidth}
\includegraphics[width=0.89\textwidth]{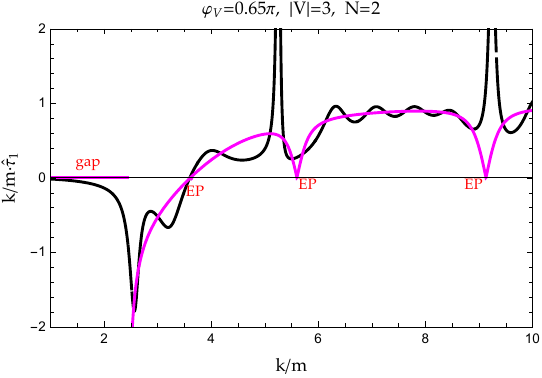}
\caption{  }\label{tau1plot2}
\end{subfigure} 
\begin{subfigure}[b]{0.38\textwidth}
\includegraphics[width=0.89\textwidth]{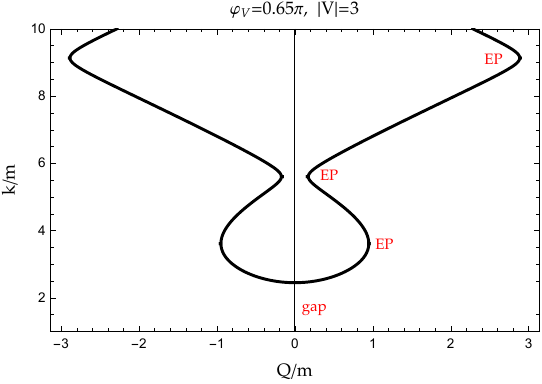}
\caption{  }\label{bandplot2}
\end{subfigure}
\caption{     (a \& c)   Comparison of  $\frac{ k}{m} \widehat{\tau}_1$   (solid black)  together with $\frac{ k}{m} \frac{d Re[Q]}{d E} $  (solid purple/light grey)  in unbroken and broken   $\mathcal{P}\mathcal{T}$-symmetric phase;   (b \& d)  The corresponding band structure plot  in unbroken and broken   $\mathcal{P}\mathcal{T}$-symmetric phase. The rest of parameters are taken as:    $m L=1$ and $m a =0.2$, where $|V|$ is dimensionless. }\label{tauplot2}
\end{figure*}

The location of moving poles  are controlled by model parameters. The spectral singularities only occur in certain range of model parameters, the boundary of the range of model parameters hence separate normal behaviour of tunneling time  and Faraday/Kerr rotation angles where $\tau_1$ and $\theta_1^{T/R}$ always remain positive   from  anomalous behaviours  where  $\tau_1$ and $\theta_1^{T/R}$  may turn negative near location of spectral singularities. When the model parameters are varied continuously, the  tunneling time  and Faraday/Kerr rotation angles in  $\mathcal{P}\mathcal{T}$-symmetric systems hence experience a phase transition-like transformation.   Using expressions in Eq.(\ref{t0r0expression}) and Eq.(\ref{RIQexpression}) as a simple example,  the conditions for spectral singularities are given by  considering $1/t_0(k) =0$: 
 \begin{equation}
     \left ( -  \frac{\cot (kL) |\sin (kL)|}{\sqrt{2}}  ,   \frac{k/m}{\sqrt{2}  |\sin (kL) | }    \right )    =   \left (  \cos \varphi_V    ,       |V|    \right )  . \label{specsingular}
\end{equation}
 The solutions of spectral singularities   can be visualized graphically by observing the intersection of a curve and a line with  $(x,y)$ coordinates given by both sides of  Eq.(\ref{specsingular}) for a fixed $ |V|$, see Fig.~\ref{spectimeplot} as a example.    For a fixed $ |V|=4$, the solutions of spectral singularities can only be found in a finite range:  $\varphi_V \in [0.4\pi ,  0.67\pi ]$, in which  the poles appear in physical sheet,  anomalous behaviour of tunneling time and Faraday/Kerr rotation angles occur and   $\tau_1$ and $\theta_1^{T/R}$ may turn negative. For a fixed  $ |V|=4$, only a single pole solution can be found in complex $k$-plane, the motion of pole  as   $\varphi_V$ is increased  is illustrated in Fig.\ref{motionpolesplot}.

For a large $N$ system, the situation is even more interesting,   the band structure and EPs     start getting involved, competing with poles and playing the roles in turning $\tau_1$ and $\theta_1^{T/R}$, see detailed discussion in Ref.~\cite{Guo:2022jyk}.     The band structure of system is clearly visible for even small size systems. The number of poles grow drastically with size, and the distribution of poles split into bands. When the poles show up inside an allowed band of system and all move across the real axis, they tend to flip the sign of entire band. In some bands where  two bands start merging together at exceptional point, the exceptional points tend to force $\tau_1$ and $\theta_1^{T/R}$ approaching zero and starts to competing with poles, so the $\mathcal{P}\mathcal{T}$-symmetric  systems become almost transparent near EPs. The fate of $\tau_1$  and $\theta_1^{T/R}$ near EPs now is  the result of two competing forces: the poles and EPs.

\subsection{Large  $N$  limit}

As number of cells is increased, all traversal time  $\tau_{1,2}$ and  FR and KR angles demonstrate fast oscillating behaviour due to $\sin ( \beta (2N+1)\Lambda )$ and $\cos ( \beta (2N+1)\Lambda )$ functions in transmission and reflection amplitudes.  For the large $N$ systems, we can introduce the traversal time   per unit cell and  FR and KR angles per unit cell, such as
\begin{align}
\hat{\tau}_{1,2}  = \frac{ \tau_{1,2}}{(2 N+1) \Lambda} .
\end{align}
The   $N \rightarrow \infty$ limit may be approached by adding a small imaginary part to $\beta$: $\beta \rightarrow \beta + i \epsilon$, where $\epsilon \gg \frac{1}{(2N+1) \Lambda}$.   As discussed in Ref.~\cite{Guo:2022jyk},  adding a small imaginary part to $\beta$ is justified by considering the averaged FR and KR angles per unit cell, which ultimately smooth out the fast oscillating behaviour of $\tau_{1,2}$ and FR and KR angles.
Therefore, as $N \rightarrow \infty$, the two components of traversal time per unit cell  approach
\begin{equation}
\hat{\tau}_1   - i \hat{\tau}_{2}  \stackrel{ N \rightarrow \infty}{\rightarrow}   \frac{d  \beta }{d E}  , \label{largeNtime}
\end{equation}
and FR and KR angles per unit cell approach
\begin{equation}
\hat{\theta}^{T}_1   - i \hat{\theta}^{T}_{2}  \stackrel{ N \rightarrow \infty}{\rightarrow} 
  \frac{g}{2n} \frac{\partial  \beta }{\partial n} ,    \ \ \ \ \ \ \ \   \hat{\theta}^{R^{(r/l)}}_{2}  \stackrel{ N \rightarrow \infty}{\rightarrow} 0. \label{largeNFR}
\end{equation}

The examples of tunneling time per unit cell for  a $\mathcal{P}\mathcal{T}$-symmetric finite system with five cells  are shown in  Fig.~\ref{tauplot2}, compared with the large $N$ limit results.  As we can see in    Fig.~\ref{tauplot2}, the $\tau_1$      oscillate around the  large $N$ limit results. Even for the small size system, we can see clearly that the   band structure of infinite periodic system is already showing up.   In broken $\mathcal{P}\mathcal{T}$-symmetric phase in   Fig.~\ref{tauplot2}, EPs can be visualized even for a small size system, where two neighbouring bands merge and the  $\mathcal{P}\mathcal{T}$ becomes totally transparent:    $\tau_{1}$   approach zero.  FR/KR angles show the similar behaviour, see e.g. Fig. 4 and Fig. 5 in Ref.\cite{Perez-Garrido:2023pfg}.

  The limiting cases  in  Eq.(\ref{largeNtime}) and  Eq.(\ref{largeNFR})  work well and  are mathematically well-defined in bands where spectral singularities are absent on real axis, the poles are all either in physical sheet or already all crossed real axis into unphysical sheet. The large $N$ limit is well defined and can be achieved by either averaging fast oscillating behaviour of $\tau_E$ or using $i \epsilon$-prescription  by shifting $k$  off real axis into complex plane.  However, in the bands where   the divergent singularities show up on the real axis and the band is still in the middle of transition between all positive and all negative band of $\tau_1$ and $\theta^{T/R}_1$, see e.g. Fig.8 in  Ref.~\cite{Guo:2022jyk},     Eq.(\ref{largeNtime}) and  Eq.(\ref{largeNFR}) break down,  and large $N$ limit becomes ambiguous and problematic.    The question of how to define a physically meaningful large $N$ limit in presence of spectral singularities  is still open.

\section{Summary}
We give a brief review on recent development of the generalization of tunneling time and anomalous behaviour of Faraday and Kerr rotation angles  in  $\mathcal{P}\mathcal{T}$-symmetric systems. Both phenomena
are closely related to each other, associated with a generalized density of states and 
exhibit a phase transition-like anomalous behaviour in certain range of model parameters. Anomalous behaviour of  tunneling time and   Faraday/Kerr   angles  in  $\mathcal{P}\mathcal{T}$-symmetric systems  is directly related to   the motion of poles of scattering amplitudes in energy/frequency complex plane. When poles show up in  physical sheets, the value of tunneling time  $\tau_1$  and    Faraday and Kerr rotation angles $\theta^{T/R}_1$ may turn negative,    which may be considered as anomalous phase of $\mathcal{P}\mathcal{T}$-symmetric systems. On the contrary, when all poles  remain in unphysical sheet, tunneling time and   Faraday/Kerr   angles of $\mathcal{P}\mathcal{T}$-symmetric systems behave just as normal Hermitian systems, which may be considered as normal phase of systems.  Both $\tau_1$  and  $\theta^{T/R}_1$ exhibit a strong enhancement when  the poles move close to real axis where spectral singularities occur.

\section{Acknowledgement }  
V.G., A.P-G. and E.J. would like to thank UPCT for partial financial support through "Maria Zambrano ayudas para la recualificación del sistema universitario español 2021-2023" financed by Spanish Ministry of Universities with funds "Next Generation" of EU. This research was supported in part by the National Science Foundation under Grant No. NSF PHY-1748958. 

\bibliographystyle{eplbib}

\bibliography{biblio.bib}

\end{document}